\documentclass[a4paper,11pt]{article}
\pdfoutput=1 

 \usepackage{jcappub}

\usepackage{tikz-feynhand}
\usepackage{graphicx}
\usepackage{xcolor}
\usepackage{caption,subcaption}
\usepackage{mathrsfs,mathtools}
\usepackage{physics,amssymb}
\usepackage{bm}
\usepackage{braket}
\usepackage{listings}
\usepackage{cases}
\usepackage{comment}
\usepackage{soul}
\usepackage{cancel}
\usepackage{cases}
\usepackage[utf8]{inputenc}
\usepackage{url}
\usepackage[normalem]{ulem}
\usepackage{xspace}
\usepackage{aas_macros}
\usepackage{acronym}
\usepackage{siunitx}
\usepackage{array}
\usepackage{ulem}

\hypersetup{colorlinks=true
,urlcolor=DARKBLUE
,anchorcolor=DARKBLUE
,citecolor=DARKBLUE
,filecolor=DARKBLUE
,linkcolor=DARKBLUE
,menucolor=DARKBLUE
,linktocpage=true
,pdfproducer=medialab
,pdfa=true
}

\definecolor{MONZA}{HTML}{CF000F}
\definecolor{DARKBLUE}{HTML}{00008b}
\definecolor{DARKMAGENTA}{HTML}{8b008b}

\newcommand{\HH}{\mathcal{H}}
\newcommand{\PP}{\mathcal{P}}

\newcommand{\ee}{\mathrm{e}}

\newcommand{\GW}{\mathrm{GW}}
\newcommand{\RD}{\mathrm{RD}}
\newcommand{\NL}{\mathrm{NL}}

\newcommand{\dk}{\frac{\dd^3k}{(2\pi)^3}}
\newcommand{\iq}[1]{\int\frac{\dd^3q_{#1}}{(2\pi)^3}}

\newcommand{\hk}[2]{h^{#2}_{\lambda^{#1}}(\tau,\bfk^{#1})}
\newcommand{\Q}[2]{Q_{\lambda_{#1}}(\bfk_{#1},\bfq_{#2})}

\acrodef{CMB}{cosmic microwave background}
\acrodef{PBH}{primordial black hole}
\acrodef{PDF}{probability density function}
\acrodef{EoM}{equation of motion}
\acrodef{GW}{gravitational wave}
\acrodef{RD}{radiation-dominated}
\acrodef{DM}{dark matter}
\acrodef{PTA}{pulsar timing array}
\acrodef{SGWB}{stochastic gravitational wave background}
\acrodef{CBC}{compact binary coalescence}
\acrodef{CL}{credible level}
\acrodef{LIGO}{Laser Interferometer Gravitational Wave Observatory}
\acrodef{LVK}{LIGO-Virgo-KAGRA}
\acrodef{SIGW}{scalar-induced gravitational wave}

\newcommand{\calI}{\mathcal{I}}

\newcommand{\bfk}{\mathbf{k}}

\newcommand{\bfq}{\mathbf{q}}

\newcommand{\ur}{\mathrm{r}}

\newcommand{\bfx}{\mathbf{x}}
\newcommand{\bmTheta}{\bm{\Theta}}
\newcommand{\CBC}{\mathrm{CBC}}

\newcommand{\mathref}{\mathrm{ref}}

\newcommand{\bae}[1]{\begin{align} #1 \end{align}}
\newcommand{\beae}[1]{\begin{equation}\begin{aligned} #1 \end{aligned}\end{equation}}

\newcommand{\bme}[1]{\begin{multline} #1 \end{multline}}

\newcommand{\bfe}[4]{
\begin{figure}[t] 
	\centering
	\includegraphics[#1]{#2}
	\caption{#3}
	\label{#4}
\end{figure}}

\newcommand{\lr}[1]{\left( #1 \right)}
\newcommand{\bce}[1]{\begin{cases} #1 \end{cases}}

\newcommand{\Real}[1]{\mathrm{Re}\left[#1\right]}

\newcommand{\Mpc}{\mathrm{Mpc}}

\newcommand{\Hz}{\mathrm{Hz}}

\begin{document}

\author[a]{Ryoto Inui,}
\author[b]{Santiago Jaraba,}
\author[a,b]{Sachiko Kuroyanagi,}
\author[c,a,d]{Shuichiro Yokoyama}

\affiliation[a]{Department of Physics, Nagoya University,\\ Furo-cho Chikusa-ku, Nagoya 464-8602, Japan}
\affiliation[b]{Instituto de F\'isica Te\'orica UAM-CSIC, \\Universidad Aut\'onoma de Madrid, 28049 Madrid, Spain}
\affiliation[c]{Kobayashi Maskawa Institute, Nagoya University,\\ Furo-cho Chikusa-ku, Nagoya 464-8602, Japan}
\affiliation[d]{Kavli IPMU (WPI), UTIAS, The University of Tokyo,\\ 5-1-5 Kashiwanoha, Kashiwa, Chiba 277-8583, Japan}

\emailAdd{inui.ryoto.a3@s.mail.nagoya-u.ac.jp}
\emailAdd{santiago.jaraba@uam.es}
\emailAdd{sachiko.kuroyanagi@csic.es}
\emailAdd{shu@kmi.nagoya-u.ac.jp}

\title{\boldmath Constraints on Non-Gaussian primordial curvature perturbation from the LIGO-Virgo-KAGRA third observing run}
\date{\today}

\abstract{The \acp{SIGW}, arising from large amplitude primordial density fluctuations, provide a unique observational test for directly probing the epoch of inflation. In this work, we provide constraints on the \ac{SIGW} background by taking into account the non-Gaussianity in the primordial density fluctuations, using the first three observing runs (O1-O3) data of the LIGO-Virgo-KAGRA collaboration. We find that the non-Gaussianity gives a non-negligible effect on the GW energy density spectrum and starts to affect the analysis of the O1-O3 data when the non-Gaussianity parameter is $F_{\NL} > 3.55$. Furthermore, the constraints exhibit asymptotic behavior given by $F_{\NL} A_g = \rm{const.}$ at large $F_{\NL}$ limit, where $A_g$ denotes the amplitude of the curvature perturbations. In the limit of large $F_{\NL}$, we placed a 95\% credible level upper limit $F_{\NL} A_g \leq 0.115, 0.106, 0.112$ at fixed scales of $10^{16}, 10^{16.5}, 10^{17} ~\Mpc^{-1}$, respectively. 
}

\maketitle
\flushbottom

\section{Introduction}
\label{sec:intro}

The first detection of \acp{GW} from a \ac{CBC}: GW150914 opened a new window to explore the universe~\cite{LIGOScientific:2016aoc}. Subsequent to this landmark detection, the \ac{LIGO} and Virgo collaborations have identified approximately 90 GW events from \ac{CBC}s up until the third observing run (O3). More recently, multiple Pulsar Timing Array (PTA) experiments reported possible evidence of the \acp{SGWB}, potentially originating from the superposition of GWs from supermassive black hole binary systems~\cite{NANOGrav:2023gor, Antoniadis:2023ott, Reardon:2023gzh, Xu:2023wog}. Nonetheless, ongoing discussions continue to search for other astrophysical and cosmological sources of the \acp{SGWB}
(see, e.g., Ref.~\cite{NANOGrav:2023hvm}).

With such substantial observational progress, the cosmological \acp{SGWB} have garnered significant attention as a means to explore the universe even before the \ac{CMB}. Various mechanisms for generating such SGWBs in the early universe prior to the \ac{CMB} have been extensively studied. These mechanisms include the production of primordial \acp{GW} originating from inflation and a preheating phase~\cite{Starobinsky:1979ty,Turner:1996ck, Khlebnikov:1997di,Garcia-Bellido:2007fiu,Aggarwal:2020olq}, cosmic strings~\cite{LIGOScientific:2017ikf,LIGOScientific:2021nrg,vanRemortel:2022fkb}, the bubble collisions at the first-order phase transition~\cite{Witten:1984rs,Kosowsky:1991ua,Kosowsky:1992vn,Caprini:2018mtu}, and more. One of the most interesting mechanisms is the \acfp{SIGW}, which can be generated from the primordial curvature perturbations through the tensor-scalar coupling predicted in the second-order cosmological perturbation theory (see, e.g., Refs.~\cite{Ananda:2006af, Baumann:2007zm, Saito:2008jc, Domenech:2021ztg}).

In general, it is challenging to directly probe small-scale primordial scalar fluctuations. However, by investigating such \acp{SIGW}, we can indirectly explore such primordial fluctuations even on much smaller scales than the large-scale structure in our Universe (see, e.g., Ref.~\cite{Inomata:2018epa}). Probing such small-scale primordial fluctuations should be valuable not only for testing inflationary models but also for indirectly searching for \acp{PBH}~\cite{Saito:2008jc}.

In Refs.~\cite{Kapadia:2020pnr, Romero-Rodriguez:2021aws}, 
the upper limits on the amplitude of primordial curvature perturbations have been placed by searching for \acp{SIGW} using the \acp{LVK} O2 and O3 data.\footnote{The search for \acp{SIGW} in the recent \ac{PTA} experiments has been frequently discussed (see, e.g., Refs.~\cite{NANOGrav:2023hvm, Franciolini:2023pbf, Cai:2023dls, Wang:2023ost, Liu:2023ymk, Abe:2023yrw, Jin:2023wri, Liu:2023pau}).} While the previous studies assumed a Gaussian distribution of the primordial curvature perturbations, in this paper, we investigate the impact of non-Gaussianity on the SIGW signal and provide new constraints on non-Gaussian primordial curvature perturbations using the \ac{LVK} O1-O3 data.

In fact, recent theoretical studies have shown that several inflationary models can generate enhanced primordial curvature perturbations with non-Gaussian distributions~(see, e.g., Refs.~\cite{Garcia-Bellido:2016dkw,Garcia-Bellido:2017aan,Cai:2018dkf, Atal:2019erb, Ezquiaga:2019ftu, Ragavendra:2021qdu, Pi:2021dft, Cai:2021zsp, LISACosmologyWorkingGroup:2022jok, Ezquiaga:2022qpw,Pi:2022ysn}). Such non-Gaussianity not only affects the abundance of \acp{PBH} (see, e.g., Refs.~\cite{Yoo:2019pma, Kitajima:2021fpq}) but also the spectrum of the \acp{SIGW} ~\cite{Cai:2018dig, Unal:2018yaa, Yuan:2020iwf, Adshead:2021hnm, Garcia-Saenz:2022tzu, Abe:2022xur}. Therefore, the observational constraints must be adjusted when considering such inflationary models predicting non-Gaussian primordial curvature perturbations. While the form of non-Gaussianity can vary significantly depending on the inflationary model, in this paper, as an initial approach, we employ the simplest non-linearity parameter, $F_{\rm NL}$, to characterize the primordial non-Gaussianity, and the monochromatic power spectrum for the primordial Gaussian fluctuations.

This paper is organized as follows. In Sec.~\ref{sec:inducedGW}, we give a brief review of \acp{SIGW} in the presence of the primordial non-Gaussianity, based on Refs~\cite{Kohri:2018awv, Abe:2022xur, Adshead:2021hnm}. Then, we briefly describe the parameter estimation method for a \ac{SGWB} searched by the cross-correlation analysis in Sec.~\ref{sec:search}. Then we place observational constraints using LVK O1-O3 data in Sec.~\ref{sec:result}. The conclusion is given in Sec.~\ref{sec:conc}. 

\section{Gravitational waves sourced by scalar perturbations}
\label{sec:inducedGW}

Here, we briefly review the \acp{SIGW} with primordial non-Gaussianity, based on Refs.~\cite{Adshead:2021hnm, Abe:2022xur, Kohri:2018awv}.
The perturbed metric in the conformal Newtonian gauge can be written as
\bae{
    \dd{s}^2 = a(\tau)^2\left[-(1+2\Phi)\dd{\tau}^2 + \left((1-2\Phi)\delta_{ij}+\frac{1}{2}h_{ij}\right)\dd{x}^i\dd{x}^j \right],
}
where $\tau$ represents the conformal time, $\Phi$ is the curvature perturbation in the Newtonian gauge, and $h_{ij}$ is the transverse traceless tensor perturbation. Here, we have neglected the vector perturbations and the scalar anisotropic stress.

The tensor perturbation can be expanded by the Fourier modes $\hk{}{}$ 
\bae{
     h_{ij}(\tau,\bfx)=\sum_{\lambda = +,\times}\int\dk \ee^{i\bfk\cdot\bfx}e^{\lambda}_{ij}\hk{}{},
     \label{eq:fourier_strain}
}
where $e^{\lambda}_{ij}(k)$ denotes the transverse traceless polarization tensors that satisfy $e^{\lambda}_{ij}e^{\lambda^\prime,ij}= \delta^{\lambda\lambda^\prime}$.

\subsection{Scalar induced gravitational waves}
At the second order, 
the Fourier mode of \acp{GW} follows the \ac{EoM} 
\bae{
    \hk{}{\prime\prime} + 2\HH \hk{}{\prime} +k^2\hk{}{}=4S_\lambda(\tau,\bfk)\,,
    \label{eq:h}
}
where $S_\lambda(\tau,\bfk)$ is the source term which is proportional to the quadratic terms of $\Phi$, $\HH=a(\tau)H(\tau)$ is the conformal Hubble parameter, and the prime denotes the partial derivative with respect to $\tau$. In terms of the primordial curvature perturbation in the uniform energy density slice on super-horizon scales, $\zeta$,
the source term can be expressed as
\bae{\label{eq:source}
    S_\lambda(\tau,\bfk)=\iq{}\Q{}{}f(\abs{\bfk-\bfq},q,\tau)\zeta(\bfq)\zeta(\bfk-\bfq)\,.
}
Here, the projection factor $\Q{}{}$ is given by
\bae{
    Q_\lambda(\bfk,\bfq)=e^\lambda_{ij}(\bfk)q^iq^j=\frac{q^2}{\sqrt{2}}\sin^2\theta\times\bce{
        \cos(2\psi) & (\lambda=+) \\
        \sin(2\psi) & (\lambda=\times)
    }\,,
}
where we work in spherical coordinates and set $\bfk$ in the $z$-direction. The source factor $f(p,q,\tau)$ in the \ac{RD} universe is given by
\bae{
    f(p,q,\tau)&=3\phi(p\tau)\phi(q\tau)+ \frac{\dd \phi(p\tau)}{\dd \ln p\tau} \frac{\dd \phi(q\tau)}{\dd \ln q\tau} +\pqty{\phi(p\tau)\frac{\dd \phi(q\tau)}{\dd \ln q\tau}+\frac{\dd \phi(p\tau)}{\dd \ln p\tau}\phi(q\tau)}, 
}
where $\phi (x)$ is the linear transfer function between $\Phi$ in the Newtonian gauge and $\zeta$
\bae{\label{eq: relation_phi_zeta}
    \Phi(\tau,\bfk)=\phi(k\tau)\zeta(\bfk)\,,
}
with\footnote{Here we consider the adiabatic scalar perturbations. In the case of the isocurvature perturbations, see e.g. Ref.~\cite{Domenech:2021and} about the transfer function.}
\bae{
    \phi(x)=-\frac{2}{3}\frac{9}{x^2}\pqty{\frac{\sin(x/\sqrt{3})}{x/\sqrt{3}}-\cos(x/\sqrt{3})}\,,
}
in the \ac{RD} era. The \ac{EoM} can be solved by the Green's function method and the particular solution is obtained as
\bae{\label{eq:hk}
    \hk{}{}=\frac{4}{a(\tau)}\int^\tau\dd{\tilde{\tau}}G_\bfk(\tau,\tilde{\tau})a(\tilde{\tau})S_\lambda(\tilde{\tau},\bfk),
}
where the Green's function $G_\bfk(\tau,\tilde{\tau})$ is given by
\bae{
    G_\bfk(\tau,\tilde{\tau})=\frac{\sin k(\tau-\tilde{\tau})}{k}\,.
}
Substituting Eq.~\eqref{eq:source} into Eq.~\eqref{eq:hk}, the two-point function of induced GWs is
\bme{\label{eq:hh}
    \braket{h_{\lambda_1}(\tau,\bfk_1)h_{\lambda_2}(\tau,\bfk_2)}=\iq{1}\iq{2}\Q{1}{1}\Q{2}{2} \\
    \times I_k(\abs{\bfk_1-\bfq_1},q_1,\tau)I_k(\abs{\bfk_2-\bfq_2},q_2,\tau) 
    \braket{\zeta(\bfq_1)\zeta(\bfk_1-\bfq_1)\zeta(\bfq_2)\zeta(\bfk_2-\bfq_2)},
}
with the kernel function
\bae{
    I_k(p,q,\tau) = 4\int^\tau\dd{\tilde{\tau}}G_\bfk(\tau,\tilde{\tau})\frac{a(\tilde{\tau})}{a(\tau)}f(p,q,\tilde{\tau}).
}

\subsection{Gravitational waves induced by non-Gaussian curvature perturbations}

Let us now consider \acp{SIGW} in the presence of non-Gaussianity in the primordial curvature perturbations. While various types of non-Gaussianities can be considered depending on inflationary models, in this work, we focus on the simplest and well-studied type known as quadratic local-type non-Gaussianity~\cite{Cai:2018dig, Unal:2018yaa, Adshead:2021hnm}. This type can be described as
\bae{\label{eq:ansatz}
\zeta(\bfx) = \zeta_g(\bfx) + F_{\NL} \zeta^2_g(\bfx),
}
where $F_{\NL}$ is the non-linearity parameter and $\zeta_g$ denotes the Gaussian curvature perturbation. In the following analysis, we assume that the power spectrum of the Gaussian curvature perturbation is monochromatic. Once we substitute the above ansatz (Eq.~\eqref{eq:ansatz}) for the non-Gaussian curvature perturbation into Eq.~\eqref{eq:hh}, we can obtain seven contributions in total, which have different momentum integrals~\cite{Adshead:2021hnm, Abe:2022xur}.

By introducing the tensor (dimensionless) power spectrum $\PP_{\lambda\lambda^{\prime}}(\tau,k)$ which is defined as
\bae{
    \braket{\hk{}{}\hk{\prime}{}}=(2\pi)^3 \delta^3(\bfk+\bfk^{\prime}) \frac{2 \pi^2}{k^3} \PP_{\lambda\lambda^{\prime}}(\tau,k)\,,
}
we can describe
\bae{
\PP_{\lambda\lambda^{\prime}}(\tau,k) = \sum_{n=1}^7 \PP_{\lambda \lambda}^{(n)}(\tau, k) \delta_{\lambda \lambda^\prime}\,,
}
and each contribution can be calculated as~\cite{Abe:2022xur}
\footnote{
The coefficients differ slightly from the expressions in Ref.~\cite{Abe:2022xur}, but the difference is due to the inclusion of the ``deformation" factor.
}
\bae{
\PP_{\lambda\lambda}^{(1)}(\tau,k) =
    2\,\calI_{\lambda\lambda}(\tau,\bfk\mid\bfq,\bfq\mid\bfq,\bfk-\bfq),
}
\beae{
\PP^{(2)}_{\lambda\lambda}(\tau,k) &=2^2\,(2!F_\NL)^2\calI_{\lambda\lambda}(\tau,\bfk\mid\bfq_1,\bfq_2\mid\bfq_2,\bfk-\bfq_2,\bfq_1-\bfq_2), \\
\PP^{(3)}_{\lambda\lambda}(\tau,k)&=2^2\,(2!F_\NL)^2\calI_{\lambda\lambda}(\tau,\bfk\mid\bfq_1,\bfq_2\mid\bfq_1,\bfq_2,\bfk-\bfq_1-\bfq_2), \\
\PP_{\lambda\lambda}^{(4)}(\tau,k) &=
    2^2\,\frac{(2!F_\NL)^2}{2!}\calI_{\lambda\lambda}(\tau,\bfk\mid\bfq_1,\bfq_1\mid\bfk-\bfq_1,\bfq_2,\bfq_1-\bfq_2),
}
\beae{
\PP^{(5)}_{\lambda\lambda}(\tau,k) &=2\,\frac{(2!F_\NL)^4}{(2!)^2}\calI_{\lambda\lambda}(\tau,\bfk\mid\bfq_1,\bfq_1\mid\bfq_1-\bfk+\bfq_3,\bfq_3,\bfq_2,\bfq_2-\bfq_1), \\
\PP^{(6)}_{\lambda\lambda}(\tau,k) &=2\,(2!F_\NL)^4\calI_{\lambda\lambda}(\tau,\bfk\mid\bfq_1,\bfq_2\mid\bfq_1-\bfq_3,\bfq_2-\bfq_3,\bfq_3,\bfq_3-\bfk), \\
\PP^{(7)}_{\lambda\lambda}(\tau,k) &=(2!F_\NL)^4\calI_{\lambda\lambda}(\tau,\bfk\mid\bfq_1,\bfq_2\mid\bfq_1-\bfk+\bfq_2-\bfq_3,\bfq_1-\bfq_3,\bfq_2-\bfq_3,\bfq_3)\,,
}
with
\bme{
    \calI_{\lambda\lambda^\prime}(\tau,\bfk\mid\bfq_1,\bfq_2\mid\bfk_1,\bfk_2,\cdots) \\ \coloneqq \frac{k^3}{2 \pi^2}
    \int\frac{\dd[3]{q_1}}{(2\pi)^3}\frac{\dd[3]{q_2}}{(2\pi)^3}\cdots Q_\lambda(\bfk,\bfq_1)Q_{\lambda^\prime}(\bfk,\bfq_2)I_k(\abs{\bfk-\bfq_1},q_1,\tau)I_k(\abs{\bfk-\bfq_2},q_2,\tau) \\
    \times P_g(k_1)P_g(k_2)\cdots.
}
Note that in the expression $\calI_{\lambda\lambda^\prime}(\tau,\bfk\mid\bfq_1,\bfq_2\mid\bfk_1,\bfk_2,\cdots)$, $\bfq_1$ and $\bfq_2$ are arguments of integrands $Q_\lambda$ and $Q_{\lambda'}$ respectively, and $\bfk_1,\bfk_2,\cdots$, which are given as the functions of $\bfq_1, \bfq_2$, are arguments of $P_g$.
Here, $P_g(k)$ is the power spectrum of the Gaussian curvature perturbation $\zeta_g$ defined as
\bae{
\braket{\zeta_g(\bfk)\zeta_g(\bfk^{\prime})} = (2\pi)^3 \delta^3 (\bfk + \bfk^{\prime}) P_g (k),
}
and for the monochromatic case, it can be described as
\bae{
P_g(k) = \frac{2 \pi^2}{k^3} A_g \delta(\ln{k} - \ln{k_*})\,,
}
where $A_g$ is the power spectrum amplitude and $k_*$ is the peak scale.
Note that the constant parameters $A_g$ and $F_{\NL}$ can be separated out from the integrals in $\calI_{\lambda\lambda^\prime}$. Then, we can easily see the order of each contribution as $\PP_{\lambda\lambda}^{(1)} = O(F_{\NL}^0 A_g^2)$, $\PP_{\lambda\lambda}^{(2,3,4)}=O(F_{\NL}^2 A_g^3)$, and $\PP_{\lambda\lambda}^{(5,6,7)}=O(F_{\NL}^4 A_g^4)$.

\bfe{width=0.95\hsize}{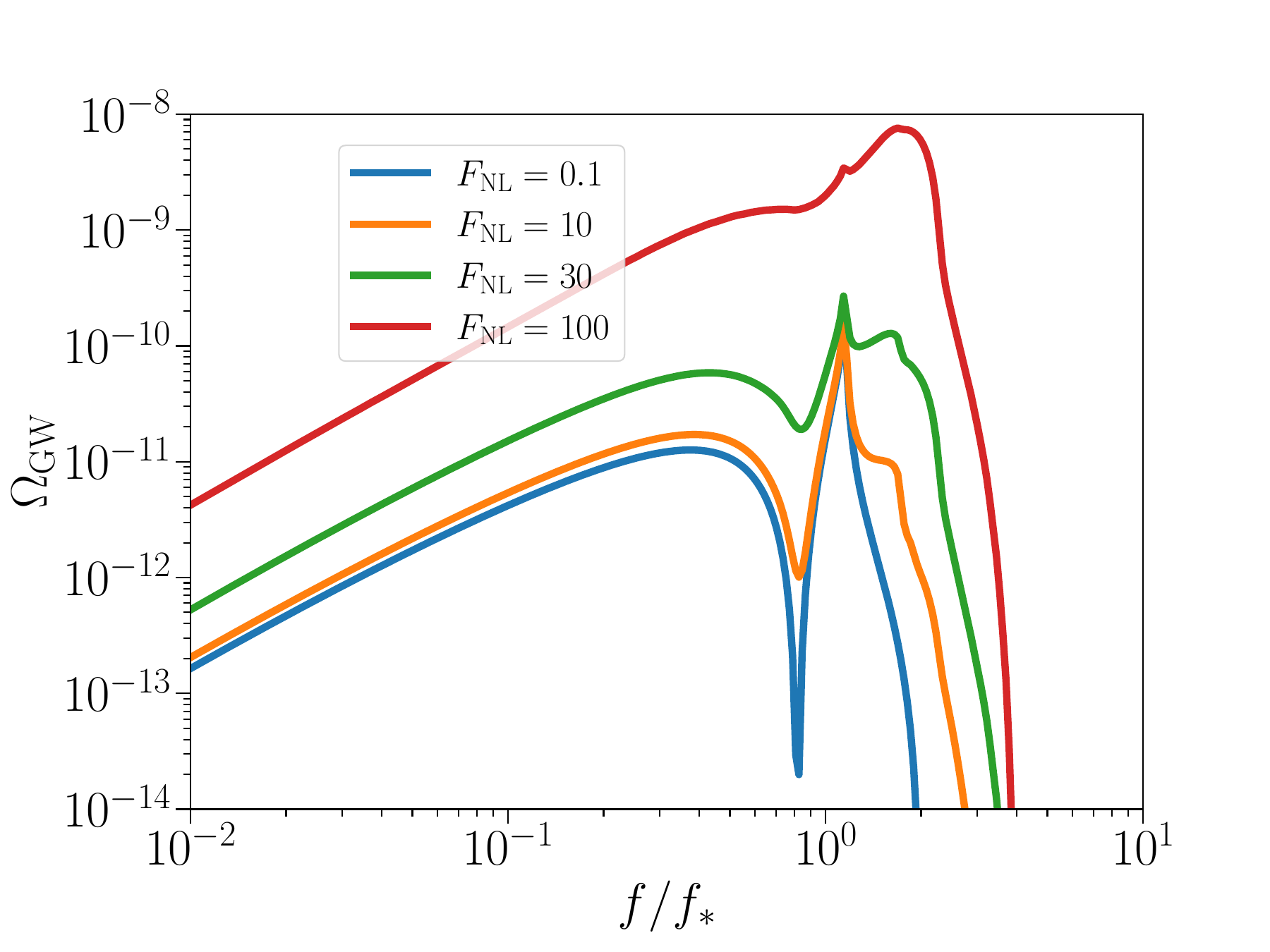}{GW energy spectrum for different $F_{\NL}$ values. We set $A_g=0.001$ as a reference value.  }{fig: GW_spectrum}

To investigate the observational constraint on the \acp{SGWB}, the density parameter per logarithmic wave number $\Omega_{\GW}\equiv\frac{1}{\rho_{\mathrm{c}}} \frac{\dd{\rho_{\GW}}}{\dd{\log{k}}}$ is commonly used, where $\rho_{\mathrm{c}}$ and $\rho_{\GW}$ respectively denote the critical energy density and the energy density of \ac{GW}. For the \acp{SIGW}, it is given by
\bae{\label{eq: OmegaGW}
\Omega_{\GW}(\tau, k) =\frac{1}{48}\lr{\frac{k}{\mathcal{H}(\tau)}}^2 
\sum_{\lambda,\lambda' = +,\times}\overline{\PP_{\lambda\lambda'}(\tau,k)}\,,
}
where the overline means the oscillation average. In the \ac{RD} era, the source term of \acp{GW} given by Eq.~\eqref{eq:source} becomes negligible soon after the curvature perturbation re-enters the horizon, and the \acp{SIGW} behave as the radiation without the source.  We denote a constant \ac{GW} density parameter in a deep subhorizon limit during the \ac{RD} era as $\Omega_\GW^\RD(k)$, and then the present density parameter can be described as (see, e.g., Ref.~\cite{Ando:2018qdb})
\bae{
\label{eq: Omg_tot}
\Omega_{\GW} (\tau_0, k)  = \Omega_{\ur,0}\frac{g^*(\tau_i)}{g^*_{,0}} \left(\frac{g^*_{s,0}}{g^*_{s}(\tau_i)} \right)^{4/3}
\Omega_\GW^\RD (k)\,,
}
where the subscript ``$0$" denotes the value at present, $\Omega_\ur$ is the density parameter of the radiation component, $g^*$ and $g^*_s$ respectively represent the effective number of relativistic degrees of freedom contributing to the energy and entropy densities, and $\tau_i$ is the conformal time when the \acp{SIGW} start to behave as radiation. Here, we assume $g^*(\tau_i) = g^*_s(\tau_i) = 106.75$, which corresponds to the epoch in which the relevant scales for \ac{LVK} band re-entered the horizon. For a more detailed description of numerically computing $\Omega_\GW^\RD (k)$, one can refer to Refs.~\cite{Adshead:2021hnm, Abe:2022xur}. Finally, in Fig.~\ref{fig: GW_spectrum}, we plot the GW energy spectrum, Eq.~(\ref{eq: Omg_tot}), for  different $F_{\NL}$ values. The relation between the gravitational waves frequency $f$ and the scale $k$ is given by
\bae{
f = \frac{k}{2\pi} =25\left(\frac{k}{1.67 \times 10^{16}~\rm Mpc^{-1}}\right) \rm Hz.
}

\section{Observational constraint by LVK detectors}
\label{sec:search}
In this section, we begin with a brief review of the detection method for \acp{SGWB} and then outline our parameter estimation setup.

\subsection{Basic formulas}
Let us consider the time series strain data of detector $I$ positioned at $\bm{x}_I$.
The total signal $s_I(t)$ can be represented by a summation of the GW signal $H_I(t)$ measured at the detector and the intrinsic noise of the detector $n_I(t)$ as 
$s_I(t) = H_I(t) + n_I(t)$ ~\cite{Allen:1997ad}. The time series strain data can be transformed in the Fourier space as
\bae{
\tilde{s}_I(f) = \tilde{H}_I(f) + \tilde{n}_I(f)\,.
 }
The measured signal $\tilde{H}_I(f)$ is described by using the Fourier amplitude of the GW signal $\tilde{h}_\lambda(f,\hat{\Omega})$ and the antenna pattern function $F^\lambda_I(f,\hat{\Omega})$, which characterizes the response to a plane wave coming from the sky position $\hat{\Omega}$, as 
\bae{
\tilde{H}_I(f) = \sum_{\lambda} \int \dd^2{\hat{\Omega}} F^\lambda_I(f,\hat{\Omega}) \tilde{h}_\lambda(f,\hat{\Omega}) e^{-i 2\pi f \bm{x}_I \cdot \hat{\Omega}}\,.
}
To extract the GW signal, the strain data are cross-correlated with a detector pair denoted as $I$ and $J$
\bae{
\label{eq: cross_signal}
\braket{\tilde{s}^*_I(f)\tilde{s}_J(f^{\prime})} = \braket{\tilde{H}^*_I(f)\tilde{H}_J(f^{\prime})} + \braket{\tilde{H}^*_I(f)\tilde{n}_J(f^{\prime})} + \braket{\tilde{n}^*_I(f)\tilde{H}_J(f^{\prime})}+
\braket{\tilde{n}^*_I(f)\tilde{n}_J(f^{\prime})}.
}
Assuming uncorrelated noise and no correlation between the \acp{GW} signal and the detector noise, the noise terms get suppressed
\footnote{If there is a correlated noise, the last term does not vanish. Schumann Resonances is one of the sources that induces such a correlated noise \cite{Thrane:2014yza, Meyers:2020qrb, Thrane:2013npa, Himemoto:2017gnw}. So far, the O3 data shows no evidence of the correlated magnetic noise~\cite{KAGRA:2021kbb}, and many statistical checks guarantee the stability and Gaussian nature of the noise, which is assumed in this paper.} 
while the first term remains and can be enhanced through the accumulation of observation time~\cite{Allen:1997ad}. Assuming an isotropic, Gaussian, stationary, and unpolarized \ac{SGWB} signal, the spectral density $S_h(f)$ can be defined by the ensemble average of the Fourier amplitude as \footnote{ In Sec.~\ref{sec:search}, we use the definition of the polarization tensor, $\ee^{\lambda}_{ij}\ee^{\lambda^{\prime},ij} = 2\delta^{\lambda \lambda^{\prime}}$, following the common notation of the GWs data analysis. Since $\Omega_{\rm GW}$ is defined to account for the difference in polarization definitions, we can confidently use $\Omega_{\rm GW}$ for comparing the theory and data, even in the presence of notation differences between Sec.~\ref{sec:inducedGW} and Sec.~\ref{sec:search}. } 
\bae{
\braket{\tilde{h}^*_{\lambda}(f,\hat{\Omega})\tilde{h}_{\lambda^{\prime}}(f^{\prime},\hat{\Omega}^{\prime})} = \frac{1}{2}\delta(f-f^{\prime}) \frac{5}{8\pi}S_h(f) \delta^{(2)}(\hat{\Omega},\hat{\Omega}^{\prime})\delta_{\lambda\lambda^{\prime}}~,
}
where $\delta(f-f^{\prime})$ denotes the finite-time width delta function, which typically can be treated as the ideal widthless delta function, and $\delta^{(2)}(\hat{\Omega},\hat{\Omega}^{\prime})=\delta(\psi-\psi^{\prime})\delta(\cos{\theta}-\cos{\theta^{\prime}})$ is a Dirac delta function on the two dimension sphere. Then the first term in Eq.~\eqref{eq: cross_signal} becomes
\bae{
\label{eq: cross_signal_ideal }
\braket{\tilde{H}^*_I(f)\tilde{H}_J(f^{\prime})} = \frac{1}{2} \delta(f-f^{\prime}) \gamma_{IJ}(f) S_h(f)\,.
}
The function $\gamma_{IJ}(f)$ is the overlap reduction function, which is determined by the relative position and orientation between the detectors and given by
\bae{
\gamma_{IJ}(f) = \frac{5}{8\pi}\sum_{\lambda} \int \dd^2{\hat{\Omega}} F^\lambda_I(f,\hat{\Omega})F^\lambda_J(f,\hat{\Omega}) e^{-i 2\pi f (\bm{x}_I-\bm{x}_J) \cdot \hat{\Omega}},
}
The spectral density can be related to the \acp{GW} density parameter Eq.~(\ref{eq: OmegaGW}) as
\bae{S_h(f)=\frac{3 H_0^2}{10 \pi^2}\frac{1}{f^3} \Omega_{\GW}(f)\,,}
where  $H_0$ is the Hubble parameter.

\subsection{Parameter estimation}
We use the LVK O1-O3 data~\cite{LIGOScientific:2016jlg, LIGOScientific:2019vic,KAGRA:2021kbb,LIGOScientific:2019lzm,KAGRA:2023pio} and perform the multi-baseline study summing the corresponding log-likelihoods for individual pairs of detectors with the Python package pygwb~\cite{Renzini:2023qtj,pygwb}.  The optimal cross-correlation estimator between times $t$ and $t + T$ is computed from data of detectors $I$ and $J$ as~\cite{LIGOScientific:2014sej,Meyers:2020qrb, KAGRA:2021kbb}
\bae{
\label{eq: cross_estimator}
C_{IJ}(f;t) = \frac{2}{T}\frac{\Real{\tilde{s}^*_I(f;t) \tilde{s}_J(f;t)}}{\gamma_{IJ}(f) S_0(f)},
}
where $S_0(f) = 3 H_0^2/(10 \pi^2 f^3)$  
is a factor that converts the GW strain power spectrum into the fractional energy density.
The variance of $C_{IJ}$ can be written in terms of the one-sided power spectral density of detector $I$, $P_I(f;t)$, and the frequency resolution $\Delta f$ as
\bae{
\label{eq: var_estimator}
\sigma^2_{IJ}(f;t) = \frac{1}{2 \Delta f T} \frac{P_I(f;t) P_J(f;t)}{\gamma^2_{IJ}(f) S_0^2(f)}.
}
In practice, the data is divided into short time segments of $T=192 {\rm s}$, so the estimator Eq.~(\ref{eq: cross_estimator}) and its variance Eq.~(\ref{eq: var_estimator}) is evaluated at each time bin. These segments are optimally combined through the following procedure 
\bae{
\hat{C}_{IJ}(f) &= \frac{\sum_K C_{IJ}(f; t_K) \sigma^{-2}_{IJ}(f; t_K) }{\sum_K  \sigma^{-2}_{IJ}(f; t_K)} , \\
\hat{\sigma}^2_{IJ}(f) &= \left(\sum_K  \sigma^{-2}_{IJ}(f; t_K)\right)^{-1},
}
where $K$ is the index of the time segments and runs from $1$ to $N$ for an $N$ set of time segments. 

In the following parameter estimation, we assume the Gaussian log-likelihood for each detector pair defined by~\cite{Mandic:2012pj, Meyers:2020qrb, Romero-Rodriguez:2021aws}
 \bae{
 \ln{p(\hat{C}_{IJ}|\bmTheta, \xi)} = - \sum_i \frac{\left[\hat{C}_{IJ}(f_i) - \xi \, \Omega_{\GW}(f_i, \bmTheta)\right]^2}{2 \hat{\sigma}^2_{IJ}},
 }
where the labels $I$ and $J$ run for different combinations of detectors \{H, L, V\} and $\xi$ is a factor that accounts for potential calibration uncertainties~\cite{Sun:2020wke,Sun:2021qcg,Whelan:2012ur}. In pygwb, it is modeled as an unknown factor with a positive normal distribution centered at 1 with a variance $\epsilon^2$ and then the likelihood is marginalized over analytically (see Appendix B of Ref.~\cite{Renzini:2023qtj}). The value of $\epsilon$ depends on the baseline: for the HL baseline, 0.072 for O1, 0.046 for O2, 0.094 for O3a and 0.148 for O3b; for the HV baseline, 0.089 for O3a and 0.123 for O3b; for the HL baseline, 0.081 for O3a and 0.108 for O3b. We use the \textit{dynesty} sampler~\cite{Speagle:2019ivv} with the default $\dd{\log{z}}=0.1$ criterion for convergence, as well as a sufficiently large nlive parameter (usually around 100,000) to get smoother posteriors.
In order to obtain a conservative bound, we include a contribution from the \ac{CBC} background, which is characterized by the power law spectrum $\Omega_{\CBC,0}(f) = \Omega_{\CBC}(f/f_{\mathref})^{2/3}$ with $f_{\mathref} = 25 ~\Hz$~\cite{KAGRA:2021kbb}.  The model $\Omega_{\GW}(f; \bmTheta) = \Omega_{\rm SIGW}(f) + \Omega_{\CBC,0}(f)$ contains the fitting parameters $\bmTheta = (A_g, F_{\NL}, k_*, \Omega_{\CBC})$ and we display the priors used for each variable in Table~\ref{table: param_prior}. We have selected prior ranges for the amplitude parameter $A_g$ and the peak scale $k_*$ such that the peak of the GW spectrum falls within the LIGO/Virgo sensitivity band. The range of $\Omega_{\CBC}$ is consistent with that used in the previous work~\cite{Romero-Rodriguez:2021aws}. The upper bound on the prior range of $F_{\NL}$ is chosen so that the tail of the posterior distribution can be observed. From a theoretical perspective, one might consider that $F_{\NL}=10^4$ is relatively large. However, as we will observe in the posterior distribution, the upper bound on $A_g$ decreases for larger values of $F_{\NL}$. We have confirmed that the curvature perturbation $\zeta$ itself does not exceed $1$ and the assumption of cosmological perturbation expansion remains valid within the parameter space we are exploring.

\begin{table}[h]
\centering
\begin{tabular}{| m{5em} | m{5cm} |}
\hline
Parameters & Prior\\
\hline
$\Omega_{\CBC}$ & $\text{Log-Uniform} [10^{-10},~10^{-7}]$\\[1ex]
\hline
$A_g$ & $\text{Log-Uniform} [10^{-3.5}, ~10^{1}]$\\ [1ex]
\hline
$F_{\NL}$ & $\text{Log-Uniform} [10^{-1}, ~10^4]$\\ [1ex]
\hline
$k_*/\Mpc^{-1}$  &  $\text{Log-Uniform} [10^{15.5}, ~10^{18.5}]$\\[1ex]
\hline
\end{tabular}
\caption{Prior distributions used for the parameter estimation. $\Omega_{\CBC}$ is the amplitude of the \ac{CBC} power spectrum at $25$Hz, $A_g$ is the primordial scalar power spectrum amplitude, $F_{\NL}$ is the non-Gaussian parameter of primordial scalar perturbations, and $k_*$ is the peak position of the primordial scalar power spectrum.}
\label{table: param_prior}
\end{table}

\section{Results}
\label{sec:result}

\bfe{width=0.95\hsize}{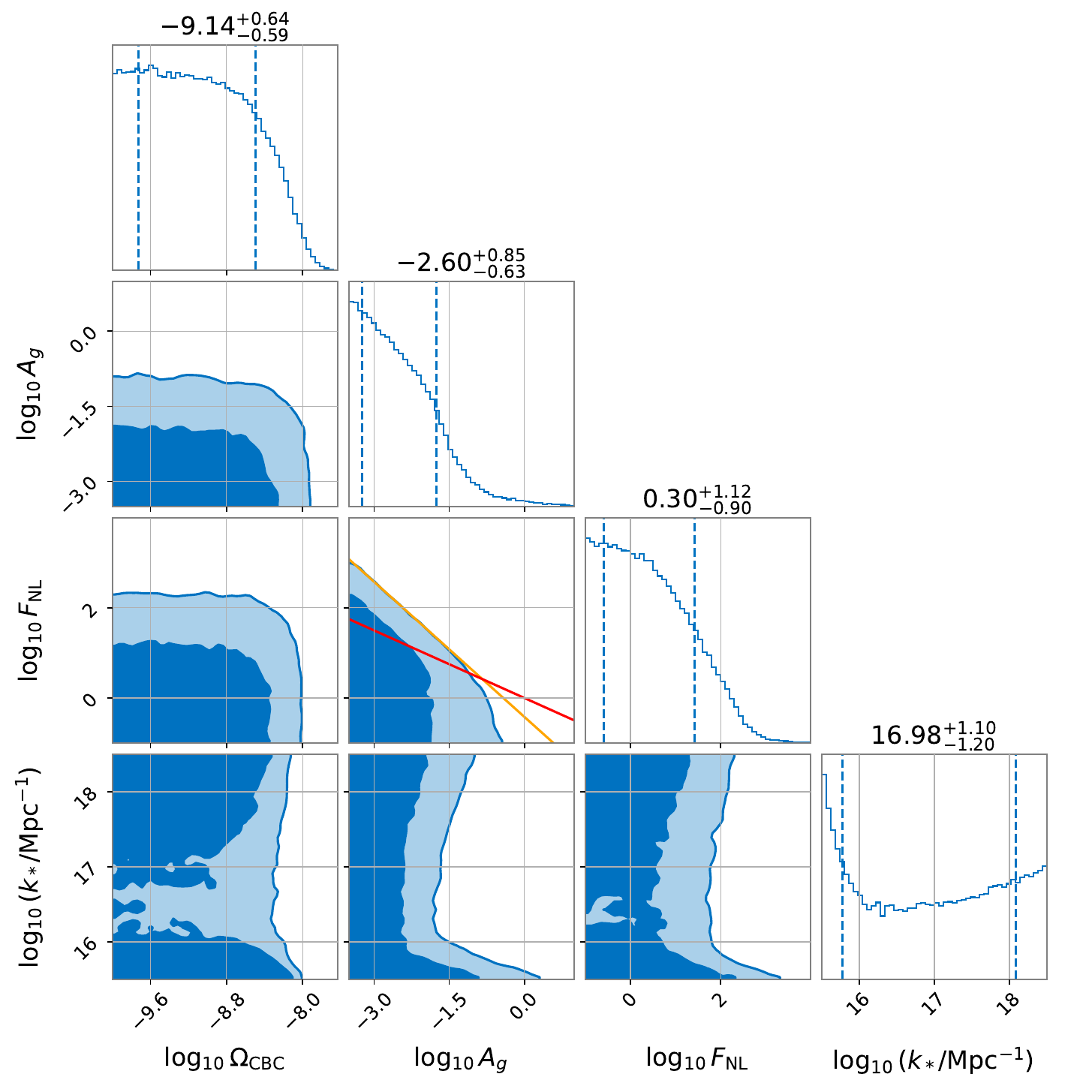}{Posterior distributions for the combined \acp{SIGW} and \ac{CBC} search. The blue and light blue contours correspond to the $68$ and $95$\% \ac{CL} allowed regions. For reference, we plot the red line corresponding to $F_{\NL}^2 A_g =1$, which is the boundary where the non-Gaussian term starts to dominate over the Gaussian contribution in the \ac{GW} power spectrum. The orange line corresponds to the asymptotic behavior at large $F_{\NL}$ values, which is characterized as $F_{\NL} A_g =0.380$.
}{fig: posterior_SGWB_CBC} 

The resultant posterior distributions with different parameters are shown in Fig.~\ref{fig: posterior_SGWB_CBC}. The orange line in the $(\log_{10}A_g-\log_{10}F_{\NL})$ plane represents the asymptotic behavior of $95$\%~\footnote{We would like to remark that the $68$ and $95$\% \ac{CL} contours do not correspond to the notion of 1 and 2$\sigma$ regions in 2 dimensions, the latter being} the default output of \textit{pygwb} provided through the \textit{corner} package~\cite{corner}. This can be a common misconception. See the \textit{corner} documentation~\cite{corner_doc} for a detailed explanation.
\ac{CL} curve characterized by 
$F_{\NL}A_g = 0.380$ 
at large $F_{\NL}$ values. 
In this regime, the dominant contribution to
$\Omega_{\rm GW}$ scales with $F_{\rm NL}^4 A^4_g$.
The red line corresponds to the boundary where non-Gaussianities begin to dominate in the \ac{SGWB} power spectrum expression. Considering that we had contributions of 
$\PP_{\lambda\lambda}^{(1)} = O(F_{\NL}^0 A_g^2)$, $\PP_{\lambda\lambda}^{(2,3,4)}=O(F_{\NL}^2 A_g^3)$, and $\PP_{\lambda\lambda}^{(5,6,7)}=O(F_{\NL}^4 A_g^4)$, we roughly set this boundary as $F^2_{NL} A_g =1$ by taking the ratio between the $\PP_{\lambda\lambda}^{(1)}$ and $\PP_{\lambda\lambda}^{(5,6,7)}$ terms. Thus, in the region above the red line, the contributions to $\Omega_{\rm GW}$ from $\PP_{\lambda\lambda}^{(5,6,7)}=O(F_{\NL}^4 A_g^4)$ are the dominant ones.
We observe a transition at $F_{\NL} \simeq 3.55$, which corresponds to the intersection between the red line and the $95$\% \ac{CL} curve. This is because, for $F_{\NL} > 3.55$, we expect the constraint is set by the $\PP_{\lambda\lambda}^{(5,6,7)}$ terms and it should follow the asymptotic behavior, which is shown by the orange line, $F_{\NL} A_g = 0.380$. On the other hand, when $F_{\NL} < 3.55$, the non-Gaussian contribution is comparable or subdominant, and thus the upper bound on $A_g$ has mild dependence on $F_{\NL}$. Note that the boundary of $F_{\rm NL} \sim 3.55$ is obtained based on the $95\%$ \ac{CL} line, and the value is specific to our choice of the \ac{CL} and the available data.

From this result, we obtain the marginalized constraint
on the primordial amplitude $A_g < 0.015$ at a $95\%$~\ac{CL} near the best sensitive band of LIGO-Virgo detectors $k_* / \Mpc^{-1} \simeq 2.04\times10^{16}$. However, we must note that this bound depends on the chosen prior range for $F_{\NL}$ and $A_g$. In general, when dealing with marginalized constraints derived from posterior distributions that do not have tails converging to zero, the constraints show a strong dependency on the selected prior range. For this reason, in Fig.~\ref{fig: Ag_k_corner}, we present the $95$\% \ac{CL} curves in the $(A_g - k_*)$ plane which have been generated in separate sampling runs by keeping the value of $F_{\NL}$ fixed. From this figure, it becomes evident that the upper bound on $A_g$ decreases for larger values of $F_{\NL}$ because it enhances the \ac{SGWB} spectral amplitude. It is also worth noting that the $k_*$ dependence of upper bound on $A_g$ varies with different values of $F_{\NL}$, and the most stringent exclusions are obtained at the scales corresponding to LIGO-Virgo's most sensitive frequency range. 
In comparison with the previous study~\cite{Romero-Rodriguez:2021aws}, although the distinction lies in our consideration of non-Gaussianity while the previous study examined a broad power spectrum, the results are consistent in the scenario of small non-Gaussianity and a narrow power spectrum.

In Fig.~\ref{fig: Ag_Fnl_corner}, we present the $95\%$~\ac{CL} curves in the $(A_g-F_{\NL})$ plane generated in different MCMC runs by fixing the peak scale $k_*$. We show three cases where $k_*/ \Mpc^{-1} =\{ 10^{16},10^{16.5}, 10^{17}\}$ and all three curves exhibit an asymptotic behavior characterized by $F_{\NL} A_g = \text{const.}$ in the limit of large $F_{\NL}$. The 95\% \ac{CL} upper limits of this quantity are
\bae{
F_{\NL}A_g \leq \left\{
\begin{array}{ll}
0.115 & ~ (k_*/\Mpc^{-1} = 10^{16}) \\
0.100 & ~ (k_*/\Mpc^{-1} = 10^{16.5}) \\
0.112 & ~ (k_*/\Mpc^{-1} = 10^{17}).
\end{array}
\right.
}
The reason why the constraint at $k_*/\text{Mpc}^{-1} = 10^{16.5}$ is stronger is simply that it exhibits a peak in the \ac{GW} spectrum within the most sensitive range of the LVK detectors. For small $F_{\NL}$, we observe different behavior between the cases of $k_*/\text{Mpc}^{-1} = 10^{16}$ and $k_*/\text{Mpc}^{-1} = 10^{16.5}$ or $10^{17}$. Below the red dashed line, the non-Gaussian contribution to the \ac{SGWB} spectrum is comparable or subdominant. As depicted in Fig.~\ref{fig: GW_spectrum}, the non-Gaussian contribution tends to appear at frequencies higher than $f_*$. In the former case, the upper bound on $A_g$ has a mild dependence on $F_{\NL}$ because the non-Gaussian effect emerges around the most sensitive frequencies of the LVK detectors. On the other hand, in the latter case, we find no dependence on $F_{\NL}$, which is because the non-Gaussian effect occurs outside the sensitive band. In this figure, we have included dotted lines representing the combinations of $F_{\NL}$ and $A_g$ that would yield a primordial black hole (PBH) abundance sufficient to constitute $100$\% of the dark matter, denoted as $f_{\rm PBH}=1$. We analytically calculated this curve using the Peak-theory~\cite{Bardeen:1985tr, Yoo:2018kvb} for estimating PBH abundance (for its application to local-type non-Gaussianity, refer to~\cite{Yoo:2019pma, Kitajima:2021fpq}). We can see that our current constraints are weaker than those imposed by the overproduction of PBHs.

\bfe{width=0.95\hsize}{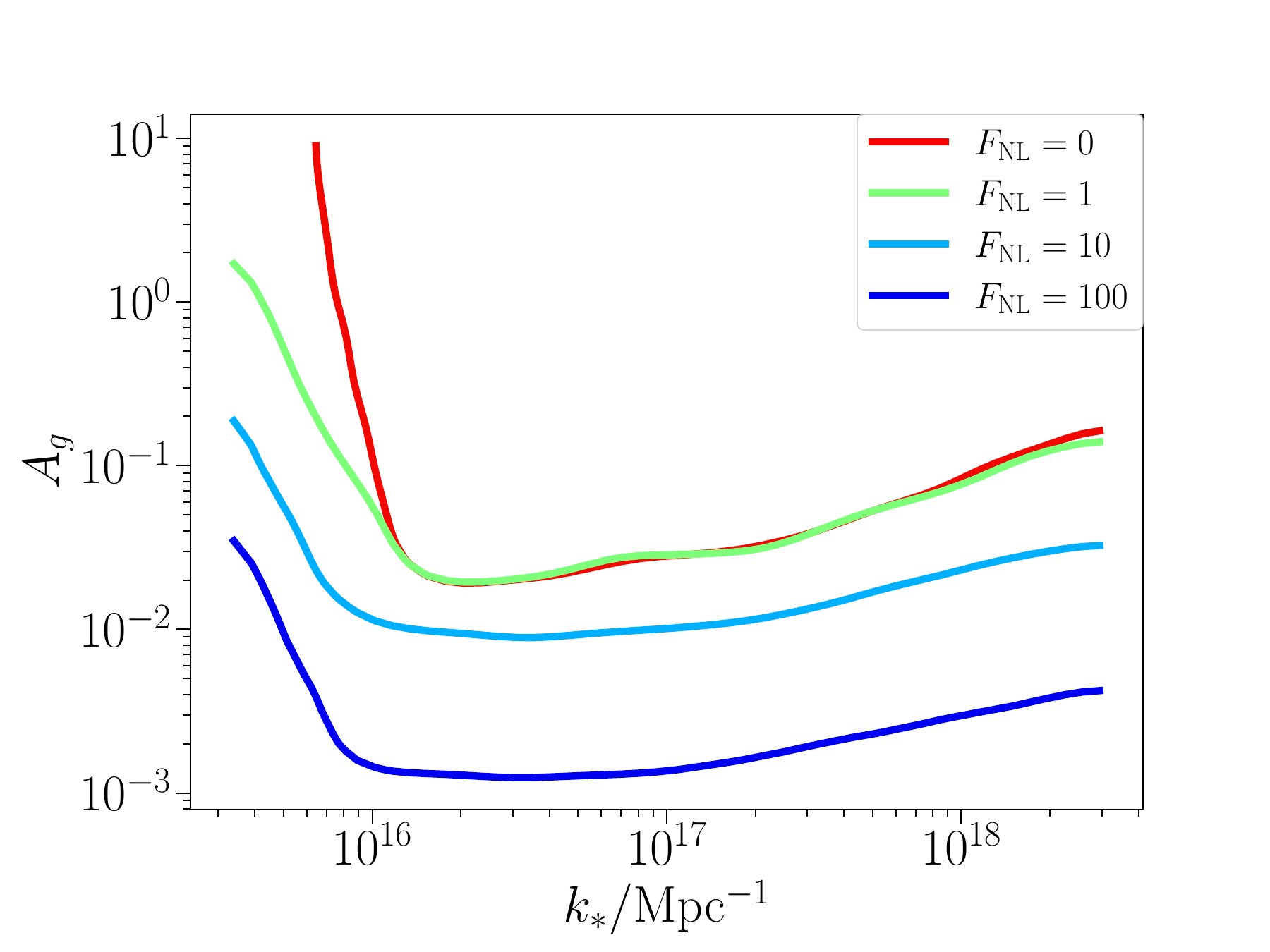}{
The 95\% \ac{CL} upper bounds on the curvature perturbation amplitude $A_g$ plotted as a function of the peak scale $k_*$. These bounds are obtained by fixing $F_{\NL}$, and different colors represent various values of $F_{\NL}$. The region below each curve corresponds to the allowed parameter space. 
}
{fig: Ag_k_corner}

\bfe{width=0.95\hsize}{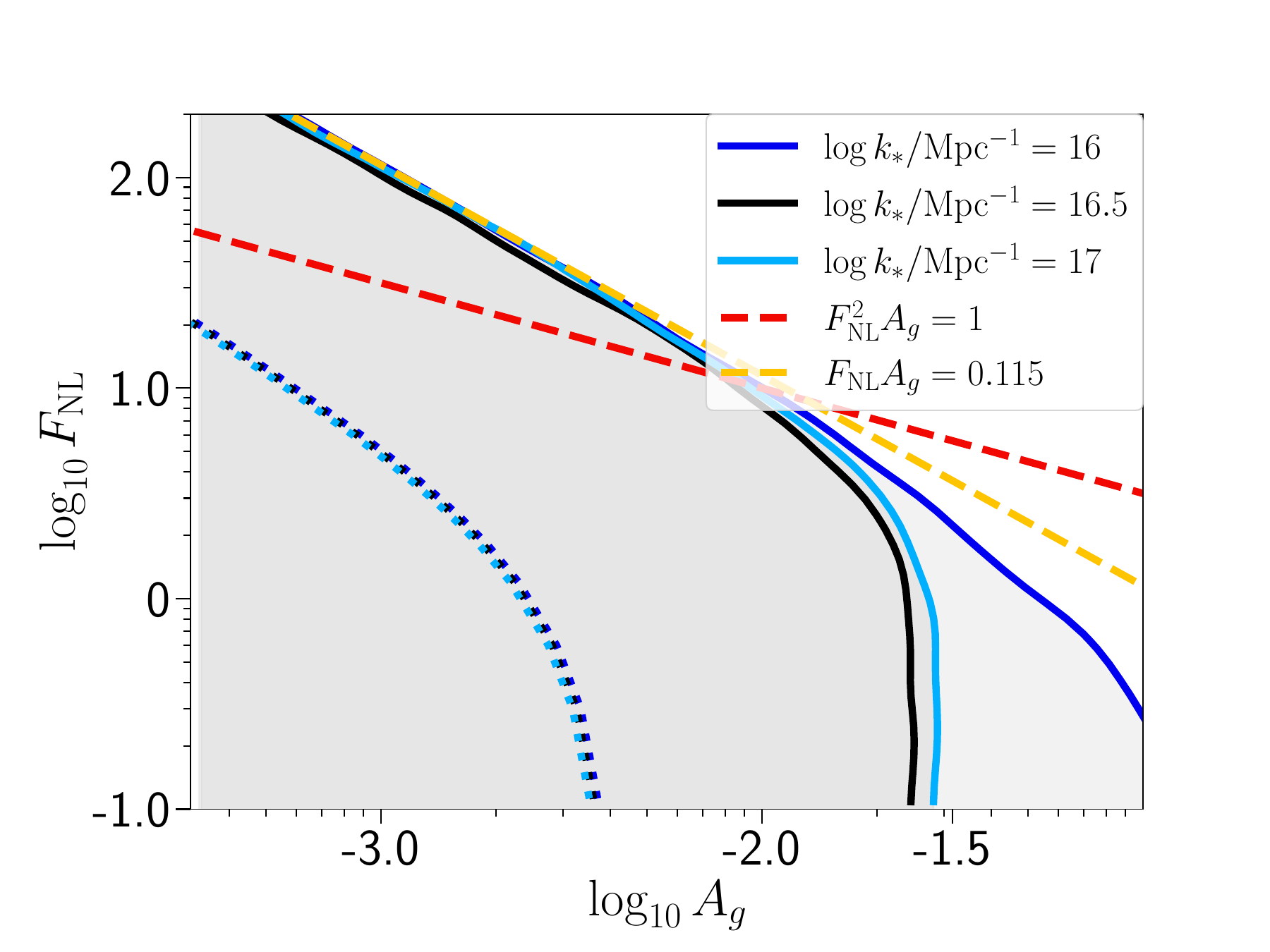}{
The $95\%$~\ac{CL} counters in the $A_g-F_{\NL}$ plane, obtained by fixing the value of the peak scale. We present three cases with $k_*/ \Mpc^{-1} =10^{16}$ (blue), $10^{16.5}$ (black), $10^{17}$ (light blue). Additionally, we include an orange line denoting the asymptotic behavior for large non-Gaussianity, where $F_{\NL} A_g = \text{const.}$, and a red line representing the boundary where the non-Gaussian term dominates over the Gaussian contribution in the GW power spectrum. Dotted lines correspond to analytically derived relations between $F_{\NL}$ and $A_g$, indicating the point at which PBHs make up 100\% of the dark matter. These lines are plotted for different values of $k_*$ but the difference is tiny.
}{fig: Ag_Fnl_corner}

\section{Discussion and Conclusion}
\label{sec:conc}

In this paper, we have investigated the constraint on the non-Gaussian primordial curvature perturbations, using the most recent data from the LVK O1-O3. By considering the second-order cosmological perturbations, the primordial curvature perturbations can induce the \acp{GW} called \acp{SIGW}, and such \acp{SIGW} are a unique probe for studying inflation physics, as the scales probed by GW interferometer experiments are much smaller than those of the \ac{CMB}.

Our main result is in Sec.~\ref{sec:result} where we performed the parameter estimation for the combined model of \acp{SIGW} and \ac{CBC}. We put an upper limit on the amplitude of the primordial curvature power spectrum $A_g$ and the strength of non-Gaussianity $F_{\NL}$. From the marginalized constraint, shown in Fig.~\ref{fig: posterior_SGWB_CBC}, the most stringent limit on the amplitude, $A_g < 0.015$ at the $95$\%~\ac{CL}, was obtained at the peak scale $k_*/\mathrm{Mpc}^{-1} \sim 2.04 \times 10^{16}$. Additionally, we observed that the influence of non-Gaussianities on the \ac{GW} power spectrum becomes non-negligible when $F_{\NL} \simeq 3.55$. Since marginalized constraints inevitably depend on the range of priors, we have also presented the analysis by fixing either the $F_{\NL}$ parameter or the peak scale $k_*$ (Figs.~\ref{fig: Ag_k_corner} and \ref{fig: Ag_Fnl_corner}). Although the current constraints appear to be weaker than those imposed by the overproduction of primordial black holes (PBHs), it is anticipated that these constraints will improve with the third-generation \ac{GW} detectors such as the Einstein Telescope (ET) and Cosmic Explorer (CE).

While we have simply assumed the quadratic order local type non-Gaussianity (the simple $F_{\NL}$ parametrization) to impose the form of non-Gaussianity, it is well known that non-Gaussian tails can vary significantly depending on the mechanisms generating large curvature perturbations. One of the primary theoretical challenges lies in developing a calculation method for the \ac{SIGW} spectrum capable of handling arbitrary shapes of non-Gaussianity. 
Once we develop a robust framework for calculating the \ac{SIGW} spectrum for all types of non-Gaussianity, we would be able to provide more valuable observational constraints for primordial curvature perturbations. This task is becoming increasingly important given the anticipated increase in observational opportunities with GWs.

\acknowledgments
\begin{sloppypar}
We acknowledge the i-LINK 2021 grant LINKA20416 by the Spanish National Research Council (CSIC), which initiated and facilitated this collaboration. The authors are grateful to Alba Romero-Rodr\'iguez for a very useful discussion. 
R.I. is supported by JST SPRING, Grant Number JPMJSP2125 and the “Interdisciplinary Frontier Next-Generation Researcher Program of the Tokai Higher Education and Research System”.
S.J. and S.K. acknowledge support from the research project PID2021-123012NB-C43 and the Spanish Research Agency (Agencia Estatal de Investigaci\'on) through the Grant IFT Centro de Excelencia Severo Ochoa No CEX2020-001007-S, funded by MCIN/AEI/10.13039/501100011033. S.J. is supported by the FPI grant PRE2019-088741 funded by MCIN/AEI/10.13039/501100011033. S.K. is supported by the Spanish Atracci\'on de Talento contract no. 2019-T1/TIC-13177 granted by Comunidad de Madrid, the I+D grant PID2020-118159GA-C42 funded by MCIN/AEI/10.13039/501100011033, the Consolidaci\'on Investigadora 2022 grant CNS2022-135211 of CSIC, and Japan Society for the Promotion of Science (JSPS) KAKENHI Grant no. 20H01899, 20H05853, and 23H00110.
S. Y. is supported by JSPS KAKENHI Grant no. 20K03968, and 23H00108.
The MCMCs have been run in the Hydra HPC cluster at the IFT. This research has made use of data or software obtained from the Gravitational Wave Open Science Center (gwosc.org), a service of the LIGO Scientific Collaboration, the Virgo Collaboration, and KAGRA. This material is based upon work supported by NSF's LIGO Laboratory which is a major facility fully funded by the National Science Foundation, as well as the Science and Technology Facilities Council (STFC) of the United Kingdom, the Max-Planck-Society (MPS), and the State of Niedersachsen/Germany for support of the construction of Advanced LIGO and construction and operation of the GEO600 detector. Additional support for Advanced LIGO was provided by the Australian Research Council. Virgo is funded, through the European Gravitational Observatory (EGO), by the French Centre National de Recherche Scientifique (CNRS), the Italian Istituto Nazionale di Fisica Nucleare (INFN) and the Dutch Nikhef, with contributions by institutions from Belgium, Germany, Greece, Hungary, Ireland, Japan, Monaco, Poland, Portugal, Spain. KAGRA is supported by the Ministry of Education, Culture, Sports, Science and Technology (MEXT), Japan Society for the Promotion of Science (JSPS) in Japan; National Research Foundation (NRF) and Ministry of Science and ICT (MSIT) in Korea; Academia Sinica (AS) and National Science and Technology Council (NSTC) in Taiwan.
\end{sloppypar}

\bibliography{main}

\providecommand{\href}[2]{#2}\begingroup\raggedright\begin{thebibliography}{10}

\bibitem{LIGOScientific:2016aoc}
{\scshape LIGO Scientific, Virgo} collaboration, \emph{{Observation of Gravitational Waves from a Binary Black Hole Merger}}, \href{https://doi.org/10.1103/PhysRevLett.116.061102}{\emph{Phys. Rev. Lett.} {\bfseries 116} (2016) 061102} [\href{https://arxiv.org/abs/1602.03837}{{\ttfamily 1602.03837}}].

\bibitem{NANOGrav:2023gor}
{\scshape NANOGrav} collaboration, \emph{{The NANOGrav 15 yr Data Set: Evidence for a Gravitational-wave Background}}, \href{https://doi.org/10.3847/2041-8213/acdac6}{\emph{Astrophys. J. Lett.} {\bfseries 951} (2023) L8} [\href{https://arxiv.org/abs/2306.16213}{{\ttfamily 2306.16213}}].

\bibitem{Antoniadis:2023ott}
J.~Antoniadis et~al., \emph{{The second data release from the European Pulsar Timing Array III. Search for gravitational wave signals}},  \href{https://arxiv.org/abs/2306.16214}{{\ttfamily 2306.16214}}.

\bibitem{Reardon:2023gzh}
D.J.~Reardon et~al., \emph{{Search for an Isotropic Gravitational-wave Background with the Parkes Pulsar Timing Array}}, \href{https://doi.org/10.3847/2041-8213/acdd02}{\emph{Astrophys. J. Lett.} {\bfseries 951} (2023) L6} [\href{https://arxiv.org/abs/2306.16215}{{\ttfamily 2306.16215}}].

\bibitem{Xu:2023wog}
H.~Xu et~al., \emph{{Searching for the Nano-Hertz Stochastic Gravitational Wave Background with the Chinese Pulsar Timing Array Data Release I}}, \href{https://doi.org/10.1088/1674-4527/acdfa5}{\emph{Res. Astron. Astrophys.} {\bfseries 23} (2023) 075024} [\href{https://arxiv.org/abs/2306.16216}{{\ttfamily 2306.16216}}].

\bibitem{NANOGrav:2023hvm}
{\scshape NANOGrav} collaboration, \emph{{The NANOGrav 15 yr Data Set: Search for Signals from New Physics}}, \href{https://doi.org/10.3847/2041-8213/acdc91}{\emph{Astrophys. J. Lett.} {\bfseries 951} (2023) L11} [\href{https://arxiv.org/abs/2306.16219}{{\ttfamily 2306.16219}}].

\bibitem{Starobinsky:1979ty}
A.A.~Starobinsky, \emph{{Spectrum of relict gravitational radiation and the early state of the universe}}, {\emph{JETP Lett.} {\bfseries 30} (1979) 682}.

\bibitem{Turner:1996ck}
M.S.~Turner, \emph{{Detectability of inflation produced gravitational waves}}, \href{https://doi.org/10.1103/PhysRevD.55.R435}{\emph{Phys. Rev. D} {\bfseries 55} (1997) R435} [\href{https://arxiv.org/abs/astro-ph/9607066}{{\ttfamily astro-ph/9607066}}].

\bibitem{Khlebnikov:1997di}
S.Y.~Khlebnikov and I.I.~Tkachev, \emph{{Relic gravitational waves produced after preheating}}, \href{https://doi.org/10.1103/PhysRevD.56.653}{\emph{Phys. Rev. D} {\bfseries 56} (1997) 653} [\href{https://arxiv.org/abs/hep-ph/9701423}{{\ttfamily hep-ph/9701423}}].

\bibitem{Garcia-Bellido:2007fiu}
J.~Garcia-Bellido, D.G.~Figueroa and A.~Sastre, \emph{{A Gravitational Wave Background from Reheating after Hybrid Inflation}}, \href{https://doi.org/10.1103/PhysRevD.77.043517}{\emph{Phys. Rev. D} {\bfseries 77} (2008) 043517} [\href{https://arxiv.org/abs/0707.0839}{{\ttfamily 0707.0839}}].

\bibitem{Aggarwal:2020olq}
N.~Aggarwal et~al., \emph{{Challenges and opportunities of gravitational-wave searches at MHz to GHz frequencies}}, \href{https://doi.org/10.1007/s41114-021-00032-5}{\emph{Living Rev. Rel.} {\bfseries 24} (2021) 4} [\href{https://arxiv.org/abs/2011.12414}{{\ttfamily 2011.12414}}].

\bibitem{LIGOScientific:2017ikf}
{\scshape LIGO Scientific, Virgo} collaboration, \emph{{Constraints on cosmic strings using data from the first Advanced LIGO observing run}}, \href{https://doi.org/10.1103/PhysRevD.97.102002}{\emph{Phys. Rev. D} {\bfseries 97} (2018) 102002} [\href{https://arxiv.org/abs/1712.01168}{{\ttfamily 1712.01168}}].

\bibitem{LIGOScientific:2021nrg}
{\scshape LIGO Scientific, Virgo, KAGRA} collaboration, \emph{{Constraints on Cosmic Strings Using Data from the Third Advanced LIGO\textendash{}Virgo Observing Run}}, \href{https://doi.org/10.1103/PhysRevLett.126.241102}{\emph{Phys. Rev. Lett.} {\bfseries 126} (2021) 241102} [\href{https://arxiv.org/abs/2101.12248}{{\ttfamily 2101.12248}}].

\bibitem{vanRemortel:2022fkb}
N.~van Remortel, K.~Janssens and K.~Turbang, \emph{{Stochastic gravitational wave background: Methods and implications}}, \href{https://doi.org/10.1016/j.ppnp.2022.104003}{\emph{Prog. Part. Nucl. Phys.} {\bfseries 128} (2023) 104003} [\href{https://arxiv.org/abs/2210.00761}{{\ttfamily 2210.00761}}].

\bibitem{Witten:1984rs}
E.~Witten, \emph{{Cosmic Separation of Phases}}, \href{https://doi.org/10.1103/PhysRevD.30.272}{\emph{Phys. Rev. D} {\bfseries 30} (1984) 272}.

\bibitem{Kosowsky:1991ua}
A.~Kosowsky, M.S.~Turner and R.~Watkins, \emph{{Gravitational radiation from colliding vacuum bubbles}}, \href{https://doi.org/10.1103/PhysRevD.45.4514}{\emph{Phys. Rev. D} {\bfseries 45} (1992) 4514}.

\bibitem{Kosowsky:1992vn}
A.~Kosowsky and M.S.~Turner, \emph{{Gravitational radiation from colliding vacuum bubbles: envelope approximation to many bubble collisions}}, \href{https://doi.org/10.1103/PhysRevD.47.4372}{\emph{Phys. Rev. D} {\bfseries 47} (1993) 4372} [\href{https://arxiv.org/abs/astro-ph/9211004}{{\ttfamily astro-ph/9211004}}].

\bibitem{Caprini:2018mtu}
C.~Caprini and D.G.~Figueroa, \emph{{Cosmological Backgrounds of Gravitational Waves}}, \href{https://doi.org/10.1088/1361-6382/aac608}{\emph{Class. Quant. Grav.} {\bfseries 35} (2018) 163001} [\href{https://arxiv.org/abs/1801.04268}{{\ttfamily 1801.04268}}].

\bibitem{Ananda:2006af}
K.N.~Ananda, C.~Clarkson and D.~Wands, \emph{{The Cosmological gravitational wave background from primordial density perturbations}}, \href{https://doi.org/10.1103/PhysRevD.75.123518}{\emph{Phys. Rev. D} {\bfseries 75} (2007) 123518} [\href{https://arxiv.org/abs/gr-qc/0612013}{{\ttfamily gr-qc/0612013}}].

\bibitem{Baumann:2007zm}
D.~Baumann, P.J.~Steinhardt, K.~Takahashi and K.~Ichiki, \emph{{Gravitational Wave Spectrum Induced by Primordial Scalar Perturbations}}, \href{https://doi.org/10.1103/PhysRevD.76.084019}{\emph{Phys. Rev. D} {\bfseries 76} (2007) 084019} [\href{https://arxiv.org/abs/hep-th/0703290}{{\ttfamily hep-th/0703290}}].

\bibitem{Saito:2008jc}
R.~Saito and J.~Yokoyama, \emph{{Gravitational wave background as a probe of the primordial black hole abundance}}, \href{https://doi.org/10.1103/PhysRevLett.102.161101}{\emph{Phys. Rev. Lett.} {\bfseries 102} (2009) 161101} [\href{https://arxiv.org/abs/0812.4339}{{\ttfamily 0812.4339}}].

\bibitem{Domenech:2021ztg}
G.~Dom\`enech, \emph{{Scalar Induced Gravitational Waves Review}}, \href{https://doi.org/10.3390/universe7110398}{\emph{Universe} {\bfseries 7} (2021) 398} [\href{https://arxiv.org/abs/2109.01398}{{\ttfamily 2109.01398}}].

\bibitem{Inomata:2018epa}
K.~Inomata and T.~Nakama, \emph{{Gravitational waves induced by scalar perturbations as probes of the small-scale primordial spectrum}}, \href{https://doi.org/10.1103/PhysRevD.99.043511}{\emph{Phys. Rev. D} {\bfseries 99} (2019) 043511} [\href{https://arxiv.org/abs/1812.00674}{{\ttfamily 1812.00674}}].

\bibitem{Kapadia:2020pnr}
S.J.~Kapadia, K.~Lal~Pandey, T.~Suyama, S.~Kandhasamy and P.~Ajith, \emph{{Search for the Stochastic Gravitational-wave Background Induced by Primordial Curvature Perturbations in LIGO\textquoteright{}s Second Observing Run}}, \href{https://doi.org/10.3847/2041-8213/abe86e}{\emph{Astrophys. J. Lett.} {\bfseries 910} (2021) L4} [\href{https://arxiv.org/abs/2009.05514}{{\ttfamily 2009.05514}}].

\bibitem{Romero-Rodriguez:2021aws}
A.~Romero-Rodriguez, M.~Martinez, O.~Pujol\`as, M.~Sakellariadou and V.~Vaskonen, \emph{{Search for a Scalar Induced Stochastic Gravitational Wave Background in the Third LIGO-Virgo Observing Run}}, \href{https://doi.org/10.1103/PhysRevLett.128.051301}{\emph{Phys. Rev. Lett.} {\bfseries 128} (2022) 051301} [\href{https://arxiv.org/abs/2107.11660}{{\ttfamily 2107.11660}}].

\bibitem{Franciolini:2023pbf}
G.~Franciolini, A.~Iovino, Junior., V.~Vaskonen and H.~Veermae, \emph{{The recent gravitational wave observation by pulsar timing arrays and primordial black holes: the importance of non-gaussianities}},  \href{https://arxiv.org/abs/2306.17149}{{\ttfamily 2306.17149}}.

\bibitem{Cai:2023dls}
Y.-F.~Cai, X.-C.~He, X.~Ma, S.-F.~Yan and G.-W.~Yuan, \emph{{Limits on scalar-induced gravitational waves from the stochastic background by pulsar timing array observations}},  \href{https://arxiv.org/abs/2306.17822}{{\ttfamily 2306.17822}}.

\bibitem{Wang:2023ost}
S.~Wang, Z.-C.~Zhao, J.-P.~Li and Q.-H.~Zhu, \emph{{Implications of Pulsar Timing Array Data for Scalar-Induced Gravitational Waves and Primordial Black Holes: Primordial Non-Gaussianity $f_{\mathrm{NL}}$ Considered}},  \href{https://arxiv.org/abs/2307.00572}{{\ttfamily 2307.00572}}.

\bibitem{Liu:2023ymk}
L.~Liu, Z.-C.~Chen and Q.-G.~Huang, \emph{{Implications for the non-Gaussianity of curvature perturbation from pulsar timing arrays}},  \href{https://arxiv.org/abs/2307.01102}{{\ttfamily 2307.01102}}.

\bibitem{Abe:2023yrw}
K.T.~Abe and Y.~Tada, \emph{{Translating nano-Hertz gravitational wave background into primordial perturbations taking account of the cosmological QCD phase transition}},  \href{https://arxiv.org/abs/2307.01653}{{\ttfamily 2307.01653}}.

\bibitem{Jin:2023wri}
J.-H.~Jin, Z.-C.~Chen, Z.~Yi, Z.-Q.~You, L.~Liu and Y.~Wu, \emph{{Confronting sound speed resonance with pulsar timing arrays}}, \href{https://doi.org/10.1088/1475-7516/2023/09/016}{\emph{JCAP} {\bfseries 09} (2023) 016} [\href{https://arxiv.org/abs/2307.08687}{{\ttfamily 2307.08687}}].

\bibitem{Liu:2023pau}
L.~Liu, Z.-C.~Chen and Q.-G.~Huang, \emph{{Probing the equation of state of the early Universe with pulsar timing arrays}}, \href{https://doi.org/10.1088/1475-7516/2023/11/071}{\emph{JCAP} {\bfseries 11} (2023) 071} [\href{https://arxiv.org/abs/2307.14911}{{\ttfamily 2307.14911}}].

\bibitem{Garcia-Bellido:2016dkw}
J.~Garcia-Bellido, M.~Peloso and C.~Unal, \emph{{Gravitational waves at interferometer scales and primordial black holes in axion inflation}}, \href{https://doi.org/10.1088/1475-7516/2016/12/031}{\emph{JCAP} {\bfseries 12} (2016) 031} [\href{https://arxiv.org/abs/1610.03763}{{\ttfamily 1610.03763}}].

\bibitem{Garcia-Bellido:2017aan}
J.~Garcia-Bellido, M.~Peloso and C.~Unal, \emph{{Gravitational Wave signatures of inflationary models from Primordial Black Hole Dark Matter}}, \href{https://doi.org/10.1088/1475-7516/2017/09/013}{\emph{JCAP} {\bfseries 09} (2017) 013} [\href{https://arxiv.org/abs/1707.02441}{{\ttfamily 1707.02441}}].

\bibitem{Cai:2018dkf}
Y.-F.~Cai, X.~Chen, M.H.~Namjoo, M.~Sasaki, D.-G.~Wang and Z.~Wang, \emph{{Revisiting non-Gaussianity from non-attractor inflation models}}, \href{https://doi.org/10.1088/1475-7516/2018/05/012}{\emph{JCAP} {\bfseries 05} (2018) 012} [\href{https://arxiv.org/abs/1712.09998}{{\ttfamily 1712.09998}}].

\bibitem{Atal:2019erb}
V.~Atal, J.~Cid, A.~Escriv\`a and J.~Garriga, \emph{{PBH in single field inflation: the effect of shape dispersion and non-Gaussianities}}, \href{https://doi.org/10.1088/1475-7516/2020/05/022}{\emph{JCAP} {\bfseries 05} (2020) 022} [\href{https://arxiv.org/abs/1908.11357}{{\ttfamily 1908.11357}}].

\bibitem{Ezquiaga:2019ftu}
J.M.~Ezquiaga, J.~Garc\'\i{}a-Bellido and V.~Vennin, \emph{{The exponential tail of inflationary fluctuations: consequences for primordial black holes}}, \href{https://doi.org/10.1088/1475-7516/2020/03/029}{\emph{JCAP} {\bfseries 03} (2020) 029} [\href{https://arxiv.org/abs/1912.05399}{{\ttfamily 1912.05399}}].

\bibitem{Ragavendra:2021qdu}
H.V.~Ragavendra, \emph{{Accounting for scalar non-Gaussianity in secondary gravitational waves}}, \href{https://doi.org/10.1103/PhysRevD.105.063533}{\emph{Phys. Rev. D} {\bfseries 105} (2022) 063533} [\href{https://arxiv.org/abs/2108.04193}{{\ttfamily 2108.04193}}].

\bibitem{Pi:2021dft}
S.~Pi and M.~Sasaki, \emph{{Primordial Black Hole Formation in Non-Minimal Curvaton Scenario}},  \href{https://arxiv.org/abs/2112.12680}{{\ttfamily 2112.12680}}.

\bibitem{Cai:2021zsp}
Y.-F.~Cai, X.-H.~Ma, M.~Sasaki, D.-G.~Wang and Z.~Zhou, \emph{{One small step for an inflaton, one giant leap for inflation: A novel non-Gaussian tail and primordial black holes}}, \href{https://doi.org/10.1016/j.physletb.2022.137461}{\emph{Phys. Lett. B} {\bfseries 834} (2022) 137461} [\href{https://arxiv.org/abs/2112.13836}{{\ttfamily 2112.13836}}].

\bibitem{LISACosmologyWorkingGroup:2022jok}
{\scshape LISA Cosmology Working Group} collaboration, \emph{{Cosmology with the Laser Interferometer Space Antenna}}, \href{https://doi.org/10.1007/s41114-023-00045-2}{\emph{Living Rev. Rel.} {\bfseries 26} (2023) 5} [\href{https://arxiv.org/abs/2204.05434}{{\ttfamily 2204.05434}}].

\bibitem{Ezquiaga:2022qpw}
J.M.~Ezquiaga, J.~Garc\'\i{}a-Bellido and V.~Vennin, \emph{{Massive Galaxy Clusters Like El Gordo Hint at Primordial Quantum Diffusion}}, \href{https://doi.org/10.1103/PhysRevLett.130.121003}{\emph{Phys. Rev. Lett.} {\bfseries 130} (2023) 121003} [\href{https://arxiv.org/abs/2207.06317}{{\ttfamily 2207.06317}}].

\bibitem{Pi:2022ysn}
S.~Pi and M.~Sasaki, \emph{{Logarithmic Duality of the Curvature Perturbation}}, \href{https://doi.org/10.1103/PhysRevLett.131.011002}{\emph{Phys. Rev. Lett.} {\bfseries 131} (2023) 011002} [\href{https://arxiv.org/abs/2211.13932}{{\ttfamily 2211.13932}}].

\bibitem{Yoo:2019pma}
C.-M.~Yoo, J.-O.~Gong and S.~Yokoyama, \emph{{Abundance of primordial black holes with local non-Gaussianity in peak theory}}, \href{https://doi.org/10.1088/1475-7516/2019/09/033}{\emph{JCAP} {\bfseries 09} (2019) 033} [\href{https://arxiv.org/abs/1906.06790}{{\ttfamily 1906.06790}}].

\bibitem{Kitajima:2021fpq}
N.~Kitajima, Y.~Tada, S.~Yokoyama and C.-M.~Yoo, \emph{{Primordial black holes in peak theory with a non-Gaussian tail}}, \href{https://doi.org/10.1088/1475-7516/2021/10/053}{\emph{JCAP} {\bfseries 10} (2021) 053} [\href{https://arxiv.org/abs/2109.00791}{{\ttfamily 2109.00791}}].

\bibitem{Cai:2018dig}
R.-g.~Cai, S.~Pi and M.~Sasaki, \emph{{Gravitational Waves Induced by non-Gaussian Scalar Perturbations}}, \href{https://doi.org/10.1103/PhysRevLett.122.201101}{\emph{Phys. Rev. Lett.} {\bfseries 122} (2019) 201101} [\href{https://arxiv.org/abs/1810.11000}{{\ttfamily 1810.11000}}].

\bibitem{Unal:2018yaa}
C.~Unal, \emph{{Imprints of Primordial Non-Gaussianity on Gravitational Wave Spectrum}}, \href{https://doi.org/10.1103/PhysRevD.99.041301}{\emph{Phys. Rev. D} {\bfseries 99} (2019) 041301} [\href{https://arxiv.org/abs/1811.09151}{{\ttfamily 1811.09151}}].

\bibitem{Yuan:2020iwf}
C.~Yuan and Q.-G.~Huang, \emph{{Gravitational waves induced by the local-type non-Gaussian curvature perturbations}}, \href{https://doi.org/10.1016/j.physletb.2021.136606}{\emph{Phys. Lett. B} {\bfseries 821} (2021) 136606} [\href{https://arxiv.org/abs/2007.10686}{{\ttfamily 2007.10686}}].

\bibitem{Adshead:2021hnm}
P.~Adshead, K.D.~Lozanov and Z.J.~Weiner, \emph{{Non-Gaussianity and the induced gravitational wave background}}, \href{https://doi.org/10.1088/1475-7516/2021/10/080}{\emph{JCAP} {\bfseries 10} (2021) 080} [\href{https://arxiv.org/abs/2105.01659}{{\ttfamily 2105.01659}}].

\bibitem{Garcia-Saenz:2022tzu}
S.~Garcia-Saenz, L.~Pinol, S.~Renaux-Petel and D.~Werth, \emph{{No-go theorem for scalar-trispectrum-induced gravitational waves}}, \href{https://doi.org/10.1088/1475-7516/2023/03/057}{\emph{JCAP} {\bfseries 03} (2023) 057} [\href{https://arxiv.org/abs/2207.14267}{{\ttfamily 2207.14267}}].

\bibitem{Abe:2022xur}
K.T.~Abe, R.~Inui, Y.~Tada and S.~Yokoyama, \emph{{Primordial black holes and gravitational waves induced by exponential-tailed perturbations}}, \href{https://doi.org/10.1088/1475-7516/2023/05/044}{\emph{JCAP} {\bfseries 05} (2023) 044} [\href{https://arxiv.org/abs/2209.13891}{{\ttfamily 2209.13891}}].

\bibitem{Kohri:2018awv}
K.~Kohri and T.~Terada, \emph{{Semianalytic calculation of gravitational wave spectrum nonlinearly induced from primordial curvature perturbations}}, \href{https://doi.org/10.1103/PhysRevD.97.123532}{\emph{Phys. Rev. D} {\bfseries 97} (2018) 123532} [\href{https://arxiv.org/abs/1804.08577}{{\ttfamily 1804.08577}}].

\bibitem{Domenech:2021and}
G.~Dom\`enech, S.~Passaglia and S.~Renaux-Petel, \emph{{Gravitational waves from dark matter isocurvature}}, \href{https://doi.org/10.1088/1475-7516/2022/03/023}{\emph{JCAP} {\bfseries 03} (2022) 023} [\href{https://arxiv.org/abs/2112.10163}{{\ttfamily 2112.10163}}].

\bibitem{Ando:2018qdb}
K.~Ando, K.~Inomata and M.~Kawasaki, \emph{{Primordial black holes and uncertainties in the choice of the window function}}, \href{https://doi.org/10.1103/PhysRevD.97.103528}{\emph{Phys. Rev. D} {\bfseries 97} (2018) 103528} [\href{https://arxiv.org/abs/1802.06393}{{\ttfamily 1802.06393}}].

\bibitem{Allen:1997ad}
B.~Allen and J.D.~Romano, \emph{{Detecting a stochastic background of gravitational radiation: Signal processing strategies and sensitivities}}, \href{https://doi.org/10.1103/PhysRevD.59.102001}{\emph{Phys. Rev. D} {\bfseries 59} (1999) 102001} [\href{https://arxiv.org/abs/gr-qc/9710117}{{\ttfamily gr-qc/9710117}}].

\bibitem{Thrane:2014yza}
E.~Thrane, N.~Christensen, R.M.S.~Schofield and A.~Effler, \emph{{Correlated noise in networks of gravitational-wave detectors: subtraction and mitigation}}, \href{https://doi.org/10.1103/PhysRevD.90.023013}{\emph{Phys. Rev. D} {\bfseries 90} (2014) 023013} [\href{https://arxiv.org/abs/1406.2367}{{\ttfamily 1406.2367}}].

\bibitem{Meyers:2020qrb}
P.M.~Meyers, K.~Martinovic, N.~Christensen and M.~Sakellariadou, \emph{{Detecting a stochastic gravitational-wave background in the presence of correlated magnetic noise}}, \href{https://doi.org/10.1103/PhysRevD.102.102005}{\emph{Phys. Rev. D} {\bfseries 102} (2020) 102005} [\href{https://arxiv.org/abs/2008.00789}{{\ttfamily 2008.00789}}].

\bibitem{Thrane:2013npa}
E.~Thrane, N.~Christensen and R.~Schofield, \emph{{Correlated magnetic noise in global networks of gravitational-wave interferometers: observations and implications}}, \href{https://doi.org/10.1103/PhysRevD.87.123009}{\emph{Phys. Rev. D} {\bfseries 87} (2013) 123009} [\href{https://arxiv.org/abs/1303.2613}{{\ttfamily 1303.2613}}].

\bibitem{Himemoto:2017gnw}
Y.~Himemoto and A.~Taruya, \emph{{Impact of correlated magnetic noise on the detection of stochastic gravitational waves: Estimation based on a simple analytical model}}, \href{https://doi.org/10.1103/PhysRevD.96.022004}{\emph{Phys. Rev. D} {\bfseries 96} (2017) 022004} [\href{https://arxiv.org/abs/1704.07084}{{\ttfamily 1704.07084}}].

\bibitem{KAGRA:2021kbb}
{\scshape KAGRA, Virgo, LIGO Scientific} collaboration, \emph{{Upper limits on the isotropic gravitational-wave background from Advanced LIGO and Advanced Virgo\textquoteright{}s third observing run}}, \href{https://doi.org/10.1103/PhysRevD.104.022004}{\emph{Phys. Rev. D} {\bfseries 104} (2021) 022004} [\href{https://arxiv.org/abs/2101.12130}{{\ttfamily 2101.12130}}].

\bibitem{LIGOScientific:2016jlg}
{\scshape LIGO Scientific, Virgo} collaboration, \emph{{Upper Limits on the Stochastic Gravitational-Wave Background from Advanced LIGO\textquoteright{}s First Observing Run}}, \href{https://doi.org/10.1103/PhysRevLett.118.121101}{\emph{Phys. Rev. Lett.} {\bfseries 118} (2017) 121101} [\href{https://arxiv.org/abs/1612.02029}{{\ttfamily 1612.02029}}].

\bibitem{LIGOScientific:2019vic}
{\scshape LIGO Scientific, Virgo} collaboration, \emph{{Search for the isotropic stochastic background using data from Advanced LIGO\textquoteright{}s second observing run}}, \href{https://doi.org/10.1103/PhysRevD.100.061101}{\emph{Phys. Rev. D} {\bfseries 100} (2019) 061101} [\href{https://arxiv.org/abs/1903.02886}{{\ttfamily 1903.02886}}].

\bibitem{LIGOScientific:2019lzm}
{\scshape LIGO Scientific, Virgo} collaboration, \emph{{Open data from the first and second observing runs of Advanced LIGO and Advanced Virgo}}, \href{https://doi.org/10.1016/j.softx.2021.100658}{\emph{SoftwareX} {\bfseries 13} (2021) 100658} [\href{https://arxiv.org/abs/1912.11716}{{\ttfamily 1912.11716}}].

\bibitem{KAGRA:2023pio}
{\scshape KAGRA, VIRGO, LIGO Scientific} collaboration, \emph{{Open Data from the Third Observing Run of LIGO, Virgo, KAGRA, and GEO}}, \href{https://doi.org/10.3847/1538-4365/acdc9f}{\emph{Astrophys. J. Suppl.} {\bfseries 267} (2023) 29} [\href{https://arxiv.org/abs/2302.03676}{{\ttfamily 2302.03676}}].

\bibitem{Renzini:2023qtj}
A.I.~Renzini et~al., \emph{{pygwb: A Python-based Library for Gravitational-wave Background Searches}}, \href{https://doi.org/10.3847/1538-4357/acd775}{\emph{Astrophys. J.} {\bfseries 952} (2023) 25} [\href{https://arxiv.org/abs/2303.15696}{{\ttfamily 2303.15696}}].

\bibitem{pygwb}
\url{https://pypi.org/project/pygwb}.

\bibitem{LIGOScientific:2014sej}
{\scshape LIGO Scientific, VIRGO} collaboration, \emph{{Searching for stochastic gravitational waves using data from the two colocated LIGO Hanford detectors}}, \href{https://doi.org/10.1103/PhysRevD.91.022003}{\emph{Phys. Rev. D} {\bfseries 91} (2015) 022003} [\href{https://arxiv.org/abs/1410.6211}{{\ttfamily 1410.6211}}].

\bibitem{Mandic:2012pj}
V.~Mandic, E.~Thrane, S.~Giampanis and T.~Regimbau, \emph{{Parameter Estimation in Searches for the Stochastic Gravitational-Wave Background}}, \href{https://doi.org/10.1103/PhysRevLett.109.171102}{\emph{Phys. Rev. Lett.} {\bfseries 109} (2012) 171102} [\href{https://arxiv.org/abs/1209.3847}{{\ttfamily 1209.3847}}].

\bibitem{Sun:2020wke}
L.~Sun et~al., \emph{{Characterization of systematic error in Advanced LIGO calibration}}, \href{https://doi.org/10.1088/1361-6382/abb14e}{\emph{Class. Quant. Grav.} {\bfseries 37} (2020) 225008} [\href{https://arxiv.org/abs/2005.02531}{{\ttfamily 2005.02531}}].

\bibitem{Sun:2021qcg}
L.~Sun et~al., \emph{{Characterization of systematic error in Advanced LIGO calibration in the second half of O3}},  \href{https://arxiv.org/abs/2107.00129}{{\ttfamily 2107.00129}}.

\bibitem{Whelan:2012ur}
J.T.~Whelan, E.L.~Robinson, J.D.~Romano and E.H.~Thrane, \emph{{Treatment of Calibration Uncertainty in Multi-Baseline Cross-Correlation Searches for Gravitational Waves}}, \href{https://doi.org/10.1088/1742-6596/484/1/012027}{\emph{J. Phys. Conf. Ser.} {\bfseries 484} (2014) 012027} [\href{https://arxiv.org/abs/1205.3112}{{\ttfamily 1205.3112}}].

\bibitem{Speagle:2019ivv}
J.S.~Speagle, \emph{{dynesty: a dynamic nested sampling package for estimating Bayesian posteriors and evidences}}, \href{https://doi.org/10.1093/mnras/staa278}{\emph{Mon. Not. Roy. Astron. Soc.} {\bfseries 493} (2020) 3132} [\href{https://arxiv.org/abs/1904.02180}{{\ttfamily 1904.02180}}].

\bibitem{corner}
D.~Foreman-Mackey, \emph{corner.py: Scatterplot matrices in python}, \href{https://doi.org/10.21105/joss.00024}{\emph{The Journal of Open Source Software} {\bfseries 1} (2016) 24}.

\bibitem{corner_doc}
``{\it {N}ote about sigmas} in corner python package documentation.'' \url{https://corner.readthedocs.io/en/latest/pages/sigmas/}.

\bibitem{Bardeen:1985tr}
J.M.~Bardeen, J.R.~Bond, N.~Kaiser and A.S.~Szalay, \emph{{The Statistics of Peaks of Gaussian Random Fields}}, \href{https://doi.org/10.1086/164143}{\emph{Astrophys. J.} {\bfseries 304} (1986) 15}.

\bibitem{Yoo:2018kvb}
C.-M.~Yoo, T.~Harada, J.~Garriga and K.~Kohri, \emph{{Primordial black hole abundance from random Gaussian curvature perturbations and a local density threshold}}, \href{https://doi.org/10.1093/ptep/pty120}{\emph{PTEP} {\bfseries 2018} (2018) 123E01} [\href{https://arxiv.org/abs/1805.03946}{{\ttfamily 1805.03946}}].

\end{thebibliography}\endgroup
\bibliographystyle{JHEP}
\end{document}